
\magnification=1200
\hsize=6truein\vsize=8.5truein


\font\bigbf=cmbx10 scaled\magstep1

\def\mbox#1{{\leavevmode\hbox{#1}}}
\def\hspace#1{{\phantom{\mbox#1}}}

\def\rD{{\rm D}}

\def\rN{{\rm N}}
\def\rS{{\rm S}}

\def\la{\lambda}
\def\si{\sigma}
\def\det{{\rm det\,}}

\def\zf{$\zeta$--function}
\def\zfs{$\zeta$--functions}

\def\frac#1/#2{\leavevmode\kern.1em
\raise.5ex\hbox{\the\scriptfont0 #1}\kern-.1em/\kern-.15em
\lower.25ex\hbox{\the\scriptfont0 #2}}
\def\sfrac#1/#2{\leavevmode\kern.1em
\raise.5ex\hbox{\the\scriptscriptfont0 #1}\kern-.1em/\kern-.15em
\lower.25ex\hbox{\the\scriptscriptfont0 #2}}

\def\gtorder{\mathrel{\raise.3ex\hbox{$>$}\mkern-14mu
             \lower0.6ex\hbox{$\sim$}}}
\def\ltorder{\mathrel{\raise.3ex\hbox{$<$}|mkern-14mu
             \lower0.6ex\hbox{\sim$}}}

\def\semidirprod{\rlap{\ss C}\raise1pt\hbox{$\mkern.75mu\times$}}

\def\for{\lower6pt\hbox{$\Big|$}}
\def\fish{\kern-.25em{\phantom{abcde}\over \phantom{abcde}}\kern-.25em}

\def\boxit#1{\vbox{\hrule\hbox{\vrule\kern3pt
        \vbox{\kern3pt#1\kern3pt}\kern3pt\vrule}\hrule}}
\def\dalemb#1#2{{\vbox{\hrule height .#2pt
        \hbox{\vrule width.#2pt height#1pt \kern#1pt
                \vrule width.#2pt}
        \hrule height.#2pt}}}

\def\square{\mathord{\dalemb{5.9}{6}\hbox{\hskip1pt}}}


\def\noin{\noindent}

\def\al{\alpha}

\def\de{\delta}
\def\Ga{\Gamma}

\def\ka{\kappa}
\def\la{\lambda}

\def\om{\omega}
\def\si{\sigma}

\def\th{\theta}
\def\ze{\zeta}

\def\eg{{\it e.g. }}

\def\pa{\partial}

\def\gap{\vskip 20truept}

\def\sect{{\vskip 10truept\noindent}}

\def\3j#1#2#3#4#5#6{\left\lgroup\matrix{#1&#2&#3\cr#4&#5&#6\cr}
\right\rgroup}


\def\nolabels{\def\eqnlabel##1{}\def\eqlabel##1{}\def\reflabel##1{}}
\def\writelabels{\def\eqnlabel##1{%
{\escapechar=` \hfill\rlap{\hskip.09in\string##1}}}%
\def\eqlabel##1{{\escapechar=` \rlap{\hskip.09in\string##1}}}%
\def\reflabel##1{\noexpand\llap{\string\string\string##1\hskip.31in}}}
\nolabels
\global\newcount\meqno \global\meqno=1
\global\meqno=1
\def\eqnn#1{\xdef #1{(\the\meqno)}%
\global\advance\meqno by1\eqnlabel#1}
\def\eqna#1{\xdef #1##1{\hbox{$(\the\meqno##1)$}}%
\global\advance\meqno by1\eqnlabel{#1$\{\}$}}
\def\eqn#1#2{\xdef #1{(\the\meqno)}\global\advance\meqno by1%
$$#2\eqno#1\eqlabel#1$$}


\global\newcount\refno
\global\refno=1 \newwrite\reffile
\newwrite\refmac
\newlinechar=`\^^J
\def\ref#1#2{\the\refno\nref#1{#2}}
\def\nref#1#2{\xdef#1{{\bf\the\refno}} 
\ifnum\refno=1\immediate\openout\reffile=refs.tmp\fi
\immediate\write\reffile{
     \noexpand\item{[{\noexpand#1}]\ }#2\noexpand\nobreak.}
     \immediate\write\refmac{\def\noexpand#1{\the\refno}}
   \global\advance\refno by1}
\def\semi{;\hfil\noexpand\break ^^J}
\def\refn#1#2{\nref#1{#2}}

\def\anp#1#2#3{{\it Ann. Phys.} {\bf {#1}} (19{#2}) #3}

\def\cmp#1#2#3{{\it Comm. Math. Phys.} {\bf {#1}} (19{#2}) #3}
\def\cqg#1#2#3{{\it Class. Quant. Grav.} {\bf {#1}} (19{#2}) #3}

\def\jmp#1#2#3{{\it J. Math. Phys.} {\bf {#1}} (19{#2}) #3}

\def\np#1#2#3{{\it Nucl. Phys.} {\bf B{#1}} (19{#2}) #3}
\def\pl#1#2#3{{\it Phys. Lett.} {\bf {#1}B} (19{#2}) #3}

\def\pams#1#2#3{{\it Proc. Am. Math. Soc.} {\bf {#1}} (19{#2}) #3}

\refn\Osgood{B.Osgood, R.Phillips and P.Sarnak, {\it J. Funct. Anal.} {\bf
80} (1988) 148}
\refn\Branson{T.P.Branson and B.\O rsted \pams{113}{91} {669}}
\refn\Dowk{J.S.Dowker {\it Effective action in spherical domains}. {\it
Comm.Math.Phys.} to be published}
\refn\Luscher{M.L\"uscher, K.Symanzik and P.Weiss \np {173}{80}{365}}
\refn\Polyakov{A.M.Polyakov \pl {103}{81}{207}}
\refn\Obe{E.C.Obe {\it J. of Math. Analysis and Appl.} {\bf 52} (1975) 648}
\refn\Stolarsky{K.B.Stolarsky {\it Mathematika} {\bf 32} (1985) 96}
\refn\Weisberger {W.I.Weisberger \cmp{112}{87}{633}}
\refn\Klein{F.Klein {\it Vorlesungen \"uber das Ikosaeder} Teubner,
Leipzig (1884)}
\refn\Forsyth{A.R.Forsyth {\it Functions of a complex variable} Cambridge
University Press, Cambridge (1893)}
\refn\DandW{J.S.Dowker and K.Worden \cqg{9}{92}289}
\refn\Aurell{E.Aurell and P.Salomonson {\it On functional determinants of
Laplacians in polygons and simplices}, Stockholm (April 1993)}
\refn\DandS{J.S.Dowker and J.P.Schofield \jmp{31}{89}{808}}
\refn\Bukhb{L.Bukhbinder, V.P.Gusynin and P.I.Fomin {\it Sov. J. Nucl.
 Phys.} {\bf 44} (1986) 534}
\refn\Alvarez{O.Alvarez \np {216}{83}{125}}
\refn\Gilkey{T.P.Branson and P.B.Gilkey {\it Comm. Partial Diff. Equations}
{\bf 15} (1990) 245}
\refn\Kennedy{G.Kennedy, R.Critchley and J.S.Dowker \anp {125}{80}{346}}
%


\vglue 1truein
\rightline {MUTP/93/24}
\gap
\centerline {\bigbf Functional determinants on regions}
\vskip 5truept
\centerline {\bigbf of the plane and sphere}
\vskip 15truept
\centerline{J.S.Dowker}
\vskip 10 truept
\centerline {\it Department of Theoretical Physics,}
\centerline{\it The University of Manchester, Manchester, England.}
\vskip 40truept
\centerline {Abstract}
\vskip 10truept
The standard formula for the change in the effective action under a
conformal transformation is extended to the case when the boundary is
piecewise smooth. We then find the functional determinants of the scalar
Laplacian on regions of the plane obtained by stereographic projection of the
fundamental domains on an orbifolded two-sphere. Examples treated are
the sector of a disk and a circular crescent. The effective action on a
spherical cap is also determined.

\rightline{October 1993}
\vfill\eject
\vskip 10truept

\vskip 10truept
\sect{\bf 1. Introduction}

\noin Apart from its intrinsic interest, the evaluation of the functional
determinant of a differential operator (usually the Laplacian) on a manifold
has a number of applications in quantum and string theory as well as in
mathematics [\Osgood], [\Branson].

Two-dimensional manifolds are the most studied and in this paper we
continue our investigation into the orbifolded sphere. The determinant of the
scalar Laplacian on a
fundamental domain, ${\cal M}$ (a spherical triangle), was determined in a
previous paper, [\Dowk]. Here we wish to employ the conformal transformation
technique of projecting the sphere stereographically onto its equatorial plane.
The corresponding change in the determinant is given by a standard formula
due to L\"uscher, Symanzik and Weiss [\Luscher], in the first instance, and
to Polyakov [\Polyakov]. We refer to it as the LSWP
relation. This will enable the determinants on the planar projections of the
fundamental spherical domains to be found.

For the hemisphere this procedure was elaborated and employed by Weisberger
[\Weisberger] to obtain, for example,
the determinant on a disk, which would be difficult to find directly in
terms of the zeros of Bessel functions, although there is some discussion of
Bessel \zfs, [\Obe], [\Stolarsky] that might prove useful.
\sect{\bf 2. The stereographic conformal transformation}

\noin This is classic material. Let $z=x+iy$ be the coordinates on the
equatorial plane of the unit sphere with metric
\eqn\spmetric{
ds^2=4{dz d\bar z\over(1+|z|^2)^2}.
}
The metric on the plane is $dz d\bar z$. Projecting from the north pole
gives the relation with the usual angular coordinates, $z=\cot\th\exp(i\phi)$.

Imagine the sphere to be tiled according to the quotient $\rS^2/\Ga$ where
$\Ga$ is one of the extended cyclic, dihedral, tetrahedral, octahedral or
icosahedral point subgroups of O(3). This tiling will project to a
partitioning of the
plane into a finite number of regions [\Klein]. (Pictures of this can be
seen in many
places, \eg [\Forsyth].) The appearance of the partitioning depends
on the choice of projection point and can be altered by rotating
the sphere, or by applying a homography to the plane.

\sect{\bf 3. The LSWP relation. Application to the orbifolded sphere.}

\noin For classically conformally invariant theories, integrating the
conformal anomaly yields the change in the effective action under
a conformal, or equivalently, a Weyl, rescaling, $g_{ij}\rightarrow\bar
g_{ij}=\exp(-2\om)g_{ij}$.

Because the domain ${\cal M}$ has corners, it is necessary to look again,
briefly, at the derivation of the LSWP relation. For simplicity, attention is
restricted to
Dirichlet conditions. Neumann conditions are treated in the appendix. We
emphasise that the
transformations considered are conformal everywhere. Schwartz-Christoffel
transformations, as used in [\DandW] and [\Aurell] for example, are excluded.

We give our results in terms of the effective action $W$ which is related
to the functional determinant by $W=\ln\det(-\nabla^2)/2=-\ze'(0)/2$

The standard anomaly equation in two dimensions is
\eqn\basic{
\de W[\bar g]={1\over4\pi}C^{(2)}_1\big[\bar g;\de\om\big]
}
with
\eqn\fold{
C^{(2)}_1\big[g;f\big]\equiv\int_{{\cal M}}C^{(2)}_1\big(g,x\big)
f(x)\,g^{1/2}\,d^2x
}

To integrate \basic\ one needs the explicit form of \fold.
The derivation of our paper [\DandS] is repeated here. The required
expression can be obtained from the simple, integrated heat-kernel
coefficient $C^{(n)}_1\big[g;1\big]$ for the conformal
operator, $-\nabla^2+\xi(n) R$ in $n$--dimensions where
$\xi(n)=(n-2)/4(n-1)$, in the following way.

{}From the variation of the zeta function, one firstly derives the relation
\eqn\reln{
C^{(n)}_1\big[g;\de\om\big]=-{1\over n-2}\de C^{(n)}_1\big[\bar g;1\big]
|_{\om=0}
}
which is all one needs, given the precise form of
$C^{(n)}_1\big[g;1\big]$.

Calculation shows that, for an $n$--dimensional domain, ${\cal M}$,
\eqn\ceeone{
C^{(n)}_1\big[g;1\big] =
{1-6\xi\over6}\int_{\cal M} R\,dV+{1\over3}\sum_i\int_
{\pa{\cal M}_i}\ka\,dS
+{1\over6}\sum_{i<j}{\pi^2-\th_{ij}^2\over\th_{ij}}\int_{I_{ij}}dL.
}
Here $\ka$ is the extrinsic curvature of the boundary parts and
$\th_{ij}$ is the dihedral angle between the boundary
components $\pa{\cal M}_i$ and $\pa{\cal M}_j$. For simplicity we have
assumed that this angle is constant along the intersection
$I_{ij}=\pa{\cal M}_i\cap\pa{\cal M}_j$. This is so for a fundamental
domain. $dV$ is the invariant volume element on ${\cal M}$ and $dS$ and
$dL$ are the induced, invariant measures on the boundary and intersection
respectively. In two-dimensions the $I_{ij}$ are the vertices of the
domain. A vertex has a content of unity.

The difference from the usual expression is the final term so
we need only look at this. Under the conformal
scaling, $g\rightarrow\bar g$, angles are preserved and the only change
is in the contents of the $(n-2)$--dimensional intersections, $I_{ij}$.
The resulting {\it extra} contribution to the variation is thus
\eqn\extrav{
\de C^{(n)}_1\big[\bar g;1\big]
=-{n-2\over6}
\int{\pi^2-\th^2\over\th}\de\om\,dL
}
where we have symbolically absorbed the summation into the integration.
{}From \reln\ we get the generalisation of the formula in [\DandS]

\eqn\foldedcee{
C^{(n)}_1\big[g;f\big] =\left({1\over6}-\xi\right)\int Rf\,dV+
\int\!\left({1\over3}\ka f-{1\over2}(n.\pa)f\right)\,dS+
{1\over6}\int{\pi^2-\th^2\over\th} f\,dL
}
where $n$ is the inward normal. (Compare also with [\Luscher], [\Alvarez], and
[\Gilkey], but without the last term.)

The next step is to set the dimension $n$ equal to two, and then to
integrate \basic. Again we need only discuss the last term in \foldedcee\
which, for $n=2$, reduces to a sum over the vertices weighted by a
factor of $\de\om(x)$ evaluated at the vertex.

The integration over $\om$
can proceed by setting $\de\om=\om\de t$ and integrating over $t$ from 0 to 1.
The result is
$$W[\bar g,g] =
{1\over24\pi}\int\om(R+\square\om)\,dV +
{1\over12\pi}\int\om\big(\ka+{1\over2}(n.\pa)\om\big)\,dS-$$
\eqn\lswp{\hspace{HHHHHHH}
{1\over8\pi}\int(n.\pa)\om\,dS+
{1\over24\pi}\sum_k{\pi^2-\th_k^2\over\th_k}\om_k
}
where the vertices are labelled simply by $k$ and where $\om_k=\om(x_k)$.

For computational purposes it is sometimes more convenient to rewrite
\lswp\ using the conformal relations
$$g^{1/2}\square\om={1\over2}\big(\bar g^{1/2}{\bar R}-g^{1/2}R\big)$$
\eqn\confrelns{
h^{1/2}(n.\pa)\om=\bar h^{1/2}\bar\ka-h^{1/2}\ka.
}
Then
$$W_a[\bar g,g] =
{1\over48\pi}\int\om(\bar g^{1/2}\bar R+g^{1/2}R)\,d^nx +
{1\over24\pi}\!\int\!\om\big(\bar h^{1/2}\bar\ka+h^{1/2}\ka\big)\,d^{n-1}x-
$$
\eqn\lswpb{
{1\over8\pi}\int(\bar h^{1/2}\bar\ka-h^{1/2}\ka)d^{n-1}x+
{1\over24\pi}\sum_k{\pi^2-\th_k^2\over\th_k}\om_k
}
where $h$ and $\bar h$ are the induced metrics on the boundary.

For $\bar g$ we choose the metric of the plane and for $g$ that of the
sphere. We assume that the effective action, $W$, on
${\cal M}$ is known according to our previous work [\Dowk]. We are then able to
find the effective action on $\overline{\cal M}$ using
\eqn\diff{
\overline W=W[\bar g,g]+W.
}

By \spmetric\ the conformal function is
\eqn\conffunc{
\om=\ln\big({2\over1+|z|^2}\big).
}
Further, $R=2$, $\bar R=0$, and $\bar g=\bar h=1$ if planar cartesian
coordinates, $(x,y)$, are used. Fundamental domains on the orbifolded
sphere
are geodesic triangles and so $\ka=0$. Since (great) circles project to
circles, $\bar\ka$ is constant over the bounding arcs on the plane.
\sect{\bf 5. The hemisphere, cap and disk}

\noin It is instructive to reconsider the hemisphere which was discussed from
this point of view by Weisberger [\Weisberger]. If the sphere is so oriented
that ${\cal M}$ is the southern hemisphere, and the
projection is from the north pole, the projected region,
$\overline{\cal M}$, is the unit, equatorial disk.
{}From \conffunc, $\om$ vanishes on the equator and the only
nonzero contributions in \lswpb\ are from the first and third integrals
yielding $W[\bar g,g]=(\ln2)/6-1/3$ (see later). Added to the effective
action on the hemisphere, this gives the effective action on the unit disk.
In order to obtain the result for a disk of radius $a$,
we use the fact that, if the linear dimensions of a manifold be scaled by
a factor of $\la$, the effective action is increased by $-\ln\la\ze(0)$
where $\ze$ refers to the \zf\ of the original figure. This result can be
applied either to the spherical or to the planar domain and we further note
that $\ze(0)$ is preserved by the stereographic projection, equalling $1/6$
for the hemisphere. Weisberger's formula is [\Weisberger]
\eqn\diskw{
W_{\rm disk}=-\ze'_R(-1)-{1\over12}\ln2-{5\over24}-{1\over6}\ln a-{1\over4}
\ln\pi.
}

This expression can be used to determine the effective action on a
spherical cap by projecting a disk of radius $a$, centred on the origin
say, back onto the unit sphere. Now we have $\ka=(1-a^2)/2a$ and
$\bar\ka=1/a$, whence, from \lswpb,

$$
W[\bar g,g]=
{1\over6}{a^2\ln(2)+\ln(1+a^2)-a^2\over1+a^2}
$$

\eqn\tocap{
+{1\over12}\ln\big({2\over1+a^2}\big)
\big(1+{4a\ka\over(1+a^2)^2}\big)-{1\over4}
\big(1-{4a\ka\over(1+a^2)^2}\big).
}

Combined with the disk expression \diskw\ this gives the Dirichlet effective
action on a cap of area $2\pi\si=4\pi a^2/(1+a^2)$ of a unit sphere as
\eqn\wcapd{
W^\rD_{\rm cap}(\si)=
-\ze'_R(-1)-{5\over24}-{1\over4}\ln2\pi
+{1\over12}\big((3\si-1)(3-\si)+\si(2-\si)\ln (2-\si)-\ln\si\big).
}

The corresponding Neumann result turns out to be

\eqn\wcapn{
W^\rN_{\rm cap}(\si)=
-\ze'_R(-1)+{7\over24}+{1\over4}\ln2\pi
+{1\over12}\big(3\si^2-8\si+3+\si(2-\si)\ln (2-\si)+5\ln\si\big).
}

Of course it is not necessary to go through the intermediary of mapping to
the plane. One can project directly to a concentric sphere.
However, we always prefer to use unit spheres.

It should be noted that $W_{\rm cap}(\si)$ diverges as the cap
becomes very small but not as it tends to the complete sphere.
In this limit $\si=2$ with,
\eqn\limitoned{
W^\rD_{\rm cap}(2)=-\ze'_R(-1)+{5\over24}-{1\over4}\ln2\pi -{1\over12}\ln2
}
and
\eqn\limitonen{
W^\rN_{\rm cap}(2)=-\ze'_R(-1)+{5\over24}+{1\over4}\ln2\pi +{5\over12}\ln2.
}
Figures 1 and 2 show some relevant plots. The Dirichlet values exhibit minima
at $\si\approx0.09525$ and $\si\approx1.87667$.


It is interesting to consider another way of introducing nonunit disks.
If the normal to the hemisphere's flat surface makes an angle
of $\al$ with the polar axis, the projection is a disk of radius $\sec\al$
whose centre is offset from the origin by a distance of $\tan\al$. We leave it
as an exercise to perform the necessary integrations in \lswpb\ and to obtain
agreement with \diskw.

Instead of rotations of the sphere, we can use
homographies of the plane, $z\rightarrow f(z)=(az+b)/(cz+d)$.
The LSWP relation in this case is fairly standard. Both $R$
and $\bar R$ are zero and the conformal function is
\eqn\homogf{
\om=-\ln|f'(z)|.
}

For a rotation through $\al$ about a horizontal axis, $f$ is
\eqn\horizrotn{
f(z)={\cos(\al/2)\,z-\sin(\al/2)\over\sin(\al/2)\,z+\cos(\al/2)}
}
although it is not essential to use this specific form.

Substitution of the relevant geometry into \lswp\ produces the integral
\eqn\integral{
{1\over24\pi}\int_0^{2\pi}\ln(1+\sin\al\cos\th)\,\big({\cos\al\over1+\sin\al
\cos\th}+1\big)\,d\th={1\over6}\ln\cos\al.
}
This is an over-elaborate way of evaluating $\ze(0)$ on the
disk.

\sect{\bf 6. Some other fundamental domains}

\noin For a generalisation of the hemisphere, we turn to the dihedral group,
the fundamental domain of which is a digon, or lune, of angle $\al=\pi/q$.
If the
rotation axis is set along the East-West direction, the projection from the
north pole is a circular crescent and the tiling of the southern hemisphere
produces an onion-like decomposition of the equatorial disc. As a
typical example we take the crescent with sides formed by half the equator
and an arc of the circle of radius $\sec\al$ and centre
$\big(\tan\al,\,0\big)$. From \lswpb\ we find, after inserting the values
of $\om$, $\bar\ka=\cos\al$,
$$
 W[\bar g,g]={1\over24\pi}\int\om\,dV+{1\over24\pi}\int\om\bar\ka\,dS-
{1\over8\pi}\int\bar\ka\,dS
$$
$$
={1\over24\pi}\int_{\pi/2}^{3\pi/2}\int_{r(\th)}^1\ln\big({2\over1+r^2}\big)
{4rdr\,d\th\over(1+r^2)^2}\,-
$$
\eqn\crescenta{
\hspace{HHHHHHHHH}{\cos\al\over24\pi}\int_{\pi/2}^{3\pi/2}\ln
\big({2\over1+r(\th)^2}\big)
\big(r^2(\th)+r'^2(\th)\big)^{1/2}d\th-{\al\over4\pi},
}
where $r(\th)$ is the radius of the inner arc. There are no contributions
from the corners because $\om$ vanishes there.

Circle geometry gives
\eqn\polar{
r^2(\th)-2r(\th)\tan\al\cos\th-1=0
}
so
$$
r(\th)=\tan\al\cos\th+\big(1+\tan^2\al\cos^2\th\big)^{1/2}.
$$
Performing the $r$--integration and combining the two $\th$--integrands we
find, writing $r$ for $r(\th)$ and after a little algebra,

\eqn\crescentc{
W[\bar g,g]={1\over12\pi}\int_{\pi/2}^{3\pi/2}{1-r^2\over1+r^2}
\ln\big({2\over1+r^2}\big)d\th-{\al\over3\pi}.
}
It does not appear that this can be taken further analytically.

The relation
\eqn\usef{
{r^2-1\over r^2+1}= {\tan\al\cos\th\over\big(1+\tan^2\al\cos^2\th\big)^{1/2}}
}
is useful and we note that the area of the crescent is
\eqn\areac{
A_{\rm crescent}=2\al+2\tan\al-(\pi-2\al)\tan^2\al.
}

The effective action on a spherical lune was determined in [\Dowk] and can be
combined with \crescentc\ according to \diff\ to give the action on the
crescent. Some Dirichlet values are displayed in Figure 3.

The extended dihedral group is another, more easily treated case.
The fundamental domain this time is a spherical triangle
with angles $\pi/2,\,\pi/2$ and $\pi/q$. The value $q=1$ corresponds to a
quartersphere and coincides with the $q=2$ dihedral case discussed above.

With the cyclic rotation axis as the polar axis, a fundamental domain having
the south pole as the vertex of angle $\pi/q$ projects to a
sector of angle $\pi/q$ of the unit equatorial disk. $\bar\ka$ is
zero on the bounding radii of the sector and unity on the circular arc.

Looking at \lswpb\ for $W[\bar g,g]$ we see that it reduces to
$$
W[\bar g,g]={1\over24\pi}\int\om dV -{1\over8q} +{q^2-1\over24q}\ln2
$$
$$
={1\over6\pi}\int_0^{\pi/q}\int_0^1\ln\big({2\over1+r^2}\big){rdr\,
d\th\over(1+r^2)^2}
-{1\over8q} +{q^2-1\over24q}\ln2
$$
$$
={1\over12q}\big(\ln2-{1\over2}\big)-{1\over8q} +{q^2-1\over24q}\ln2
$$
\eqn\dihedrala{
=-{1\over6q}+{q^2+1\over24q}\ln2.
}
Combining this at $q=1$ with the value obtained in [\Dowk] for the
quartersphere, we find the Dirichlet effective action on half a unit disk
to be,
\eqn\halfdisk{
W_\rD={1\over24}-{1\over3}\ln2-{1\over2}\ze'_R(-1)-{1\over4}\ln\pi.
}
Figure 4 plots the values for other values of $q$.

There are many other possibilities for projection and it does not seem
worthwhile giving an exhaustive treatment.
We leave the other orbifolds for possible consideration at another time.


\noin

\sect{\bf Appendix}

\noin A zero-mode occurs for Neumann conditions and there is
the further complication that conformal invariance is lost in
$n$-dimensions. To overcome this, we consider, as in [\Kennedy],
Robin conditions
\eqn\robinc{
\big(\psi+(n.\pa)\big)\phi\big|_{\pa{\cal M}}=0,
}
where $\psi$ is a function, somewhat similar to a gauge potential, defined on
the boundary, whose conformal
behaviour is designed to compensate for terms introduced by the derivative.
This behaviour is, under $g\rightarrow\bar g$,
\eqn\confpsi{
\psi\rightarrow\bar\psi=e^{\om}\big(\psi+{1\over2}(n-2)(n.\pa)\om\big).
}

We can determine the folded coefficient $C_1^{(2)}[g;f]$ in the
Neumann case following the procedure outlined in section 3. The simple
integrated Robin coefficient in $n$-dimensions is
\eqn\ceeoner{
C^{(n)}_1\big[g;1\big] =
{1-6\xi\over6}\int_{\cal M} R\,dV+\sum_i\int_
{\pa{\cal M}_i}({1\over3}\ka-2\psi)\,dS
+{1\over6}\sum_{i<j}{\pi^2-\th_{ij}^2\over\th_{ij}}\int_{I_{ij}}dL.
}

Actually this has been derived only for smooth boundaries but since any
extra terms are due to the {\it local} influence of the intersections,
$I_{ij}$, dimensional arguments show that there can be no mixing between $\psi$
and these terms.

Because conformal invariance is being maintained, equation \reln\ is still
valid and, from \confpsi, we see that there will be a cancelling factor of
$(n-2)$ to give the required coefficient,

\eqn\foldedceer{
C^{(n)}_1\big[g;f\big] ={1-6\xi\over6}\int Rf\,dV+
\int\!\left({1\over3}\ka-2\psi +{1\over2}(n.\pa)\right)\!f\,dS+
{1\over6}\int{\pi^2-\th^2\over\th} f\,dL.
}

In two dimensions, Neumann conditions are conformally invariant. We can
then consistently set $\psi$ to zero and integrate the anomaly equation
\eqn\basicn{
\de W[\bar g]={1\over4\pi}C^{(2)}_1\big[\bar g;\de\om\big]+
{1\over2}\de\bar V
}
using \foldedceer\ to give

$$W[\bar g,g] =
{1\over2}\ln\big({\bar V\over V}\big) +
{1\over24\pi}\int\om(R+\square\om)\,dV +
$$
\eqn\lswpn{\hspace{HHHHHHH}
{1\over12\pi}\int\om\big(\ka+{1\over2}(n.\pa)\om\big)\,dS+
{1\over8\pi}\int(n.\pa)\om\,dS+
{1\over24\pi}\sum_k{\pi^2-\th_k^2\over\th_k}\om_k.
}

  \immediate\closeout\reffile
  \noindent{{\bf References}}\bigskip\frenchspacing

  \input refs.tmp\vfill\eject\nonfrenchspacing
\sect{\bf Figure Captions}
\vskip 10truept
\noin Figures 1 and 2. Effective action, W, for Dirichlet and Neumann
conditions on a spherical cap of area $2\pi\si$.
\vskip 10truept
\noin Figure 3. Effective action for Dirichlet conditions on a planar
crescent of angle $\pi/q$ and side curvatures of 1 and $\cos(\pi/q)$
\vskip 10truept
\noin Figure 4. Dirichlet effective action on a sector of angle
$\pi/q$ of a unit disk.
\vskip 10truept
\end